# Dynamic behavior of polar nanoregions in re-entrant relaxor 0.6Bi(Mg$_{1/2}$Ti$_{1/2}$)O$_3$-0.4PbTiO$_3$

Kaiyuan Chen[1], Qi Zhang[2], Jia Liu[3], Jie Wang[4], Zhencheng Lan[1], Liang Fang[1], Changzheng Hu[1], Nengneng Luo[4], Biaolin Peng[5]*, Changbai Long[5]*, Dawei Wang[3]*, Laijun Liu[1]*

[1]*Guangxi Key Laboratory of Optical and Electronic Materials and Devices, College of Materials Science and Engineering, Guilin University of Technology, Guilin, 541004, China*

[2]*BCMaterials Basque center for materials, application & nanostructures, Basque, Spain*

[3]*School of Microelectronics and State Key Laboratory for Mechanical Behaviour of Materials, Xi'an Jiaotong University, Xi'an, 710049, China*

[4]*Key Laboratory for RF Circuits and Systems, Ministry of Education; Key Laboratory of Large Scale Integrated Design, Hangzhou Dianzi University, Hangzhou, 310018 China*

[5]*School of Physical Science & Technology and Guangxi Key Laboratory for Relativistic Astrophysics, Guangxi University, Nanning, 530004, China*

[6]*School of Advanced Materials and Nanotechnology, Xidian University, Xi'an, 710071, China*

## Abstract

The existence of polar nanoregions is the most important characteristic of ferroelectric relaxors, however, the size determination and dynamic of PNRs remains uncertain. We reveal a re-entrant relaxor behavior and ferroelectric-paraelectric transition coexists in complex perovskite oxide 0.6Bi(Mg$_{1/2}$Ti$_{1/2}$)O$_3$-0.4PbTiO$_3$. Two dielectric anomalies (i) the low-temperature re-entrant relaxor transition and (ii) the high-temperature diffuse phase transition (DPT) were described by the phenomenological statistical model. The sizes of the two kinds of polar nanoregions (PNRs) corresponding to two ferroelectric states were obtained. The dynamic of PNRs were analyzed using isothermal electrical modulus, which shows three critical temperatures associated with the diffuse phase transition, the formation and freezing of PNRs, 1* Corresponding authors. E-mail addresses: blpeng@xidian.edu.cn (B. Peng); longchangbai@xidian.edu.cn (C. Long); dawei.wang@xjtu.edu.cn (D. Wang), 2009011@glut.edu.cn (L. Liu).

respectively. The temperature evolution of the PNRs evolution depends on the stoichiometry of bismuth. The results provide new insights into the dynamic behavior of PNRs and the modification way of re-entrant relaxor behavior.

**1 Introduction**

Re-entrant relaxors have long been attracting considerable attention in view of their unique physical properties.[1-3] The re-entrant relaxor behavior is characterized by a strongly frequency-dependent of the real part of permittivity, which is supplemented by a dielectric plateau for the temperatures higher than the maximum permittivity temperature ($T_m$). Moreover, further increasing the temperature leads to a diffuse phase transition (DPT) around the ferroelectric to paraelectric phase transition temperature.[4]

The term "re-entrant" was used to mean that upon cooling/heating, a highly ordered ferromagnetic/ferroelectric state emerges from a complex coexistence of relaxor and ferroelectric state with less order, which resembles cluster-glass or spin-glass phases.[5,6] The similarity between the re-entrant transition and the disordered state grown in macroscopic order phase have been studied for a long time.[2,4,5] It has been found that in many re-entrant relaxors, the relationship between the probing frequency and $T_m$ can be described by the Vogel-Fulcher law.[7] It is also interesting that the activation energy is very high (0.2-0.3) eV for re-entrant relaxor, comparing to the activation energy of 0.01-0.05 eV for canonical dipole-glass and 0.02-0.15 eV for ferroelectric relaxors. It suggests that the re-entrant relaxor behavior is a typical weakly coupled relaxor behavior and thermally activated process.[8,9] The re-entrant relaxor behavior have been reported in Pb/Bi perovskite oxides.[10,11] Chen, *et al.*[12] deemed that the re-entrant relaxor of $0.65Bi(Mg_{1/2}Ti_{1/2})O_3$-$0.35PbTiO_3$ originates from the complex ionic substitution in the A and/or B sites, suggesting that clustering of B-site ions promotes the PNRs to enters a more disordered state, where the relaxor state of $0.65Bi(Mg_{1/2}Ti_{1/2})O_3$-$0.35PbTiO_3$ is attributed to the nucleation and growth of the PNRs. The size of PNRs is dominated by the clustering of the same-type ions and/or quenched random electric fields caused by heterovalent ions.[13] However, determining or tuning the size of PNRs in either the paraelectric state or the relaxor state is very difficult.

Since PNRs appear in the polar phase in the re-entrant relaxors, their dynamics and evolution (e.g., the orientation of PNRs) with respect to temperature and stoichiometry are critical to understand the characteristic and origin of re-entrant phenomenon. In this paper, we investigated the re-entrant relaxor behavior of $0.6Bi(Mg_{1/2}Ti_{1/2})O_3$-$0.4PbTiO_3$ ceramics by using (i) the phenomenological statistical model as described in Ref. [20], (ii) the Vogel-Fulcher law, and (iii) isothermal electrical modulus. Our analysis provides key parameters to reveal the different dynamics of PNRs in the re-entrant region and the DPT region. In particular, we are able to qualitatively determine the size of PNRs in the re-entrant dipole glass-like relaxor and the DPT.

**2 Experimental procedures**

The ceramics $0.6Bi_{(1-x)}(Mg_{1/2}Ti_{1/2})O_3$-$0.4PbTiO_3$ with $x$=-0.02, 0.00, 0.02, 0.04 ($0.6B_{(1-x)}MT$-$0.4PT$), which have a perovskite structure, were prepared with high purity raw materials and sintered at 1050 °C ~ 1100 °C for 3 h in air by a two-step procedure.[14]

Room temperature crystal symmetry of the ceramics were analyzed by a PANalytical X-Pert-PRO powder X-ray diffractometer with Cu*Kα* radiation ($\lambda$=1.54059 Å). Full profile Rietveld refinements were performed using the software package FULLPROF. The peak profile shape was described by a pseudo-voigt function. For each composition, thermal etching was performed at 1000°C for 30 min. The surface microstructure of the ceramics was examined using a field emission scanning electron microscope (FE-SEM, Model S4800, Hitachi, Japan). The samples were polished and coated with silver paste on both sides for electric measurements. The dielectric properties of the ceramics were measured using an impedance analyzer (Agilent 4294A, America) connected to a tube furnace. The probing frequency ranges from 100 Hz to 1 MHz and the temperature ranges from 300 K to 960 K. The ferroelectric hysteresis loop $P(E)$ and the leakage current were measured using a ferroelectric test system (TF 2000 aixACCT Systems GmbH, Germany) with triangular voltage waveform at 1 Hz in the temperature range from 293 K to 453 K. Impedance spectroscopy was collected using an impedance analyzer (Agilent 4294A, America) from 520 K to 960 K.

## 3 Results and discussion

All the major diffraction peaks in the XRD profile can be indexed by a perovskite structure according to the X-ray diffraction (XRD) pattern of the $0.6B_{(1-x)}MT$-$0.4PT$ ceramics (Fig. 1). Unfortunately, there are impurities in the sample that give rise to some extra peaks. The reflection of the impurity $Bi_{7.68}Ti_{0.32}O_{13.16}$ (JCPDS#01-087-1897, tetragonal phase, space group: *P42/nmc*) increases with the bismuth addition, indicating more $Bi_2O_3$ dissolving from the lattice.[4] No surperlattice reflection can be observed from the diffraction profiles, suggesting the absence of A/B sites ordering. The mixture phase model *P*4*mm*+*Pm*-3*m* was employed to carry out the Rietveld refinement of the XRD pattern. The cell parameters of the sample were calculated and the cell volume and the fraction of composition in the tetragonal phase and the cubic phase are listed in Table 1. The phase fractions of the tetragonal and cubic phases are near 60% and 40%. It is noteworthy that the *Pm*-3*m* space group is an ideal perovskite structure, in which all atoms are located on sites with a center of inversion.[15] It is easy to form nano-scale entities throughout the temperature regime of common interest on a long-range pseudo-cubic symmetry as a matrix in ceramics of complex A-/B-site with electronic lone pair.[11,15] As the value of *c/a* of tetragonal ferroelectric phase is very low, 1.0178(9) ~ 1.0200(0), it is extremely difficult for the samples to form a large polarization.

To characterize the microstructure, SEM micrographs of the $0.6B_{(1-x)}MT$-0.4PT ceramics were obtained, which are shown in Fig. 2. All ceramics show dense, homogeneous grains without pores. With the increase of bismuth, there is no obvious change of the grain size. The average grain sizes of the $0.6B_{0.98}MT$-0.4PT, 0.6BMT-0.4PT, $0.6B_{1.02}MT$-0.4PT, and $0.6B_{1.04}MT$-0.4PT are 0.89 μm, 0.82 μm, 0.91 μm, and 0.83 μm, respectively. The grain size for all ceramics follows a Gaussian distribution. The appearance of liquid phase during sintering process could enhance the diffusion rate of ions across grain boundaries and/or weld the smaller grains.[16] Several larger grains might be attributed to the liquid phase sintering mechanism.

Temperature dependence of the real part of permittivity (*ε'*) and dielectric loss

(tan$\delta$) for 0.6B$_{(1-x)}$MT-0.4PT ceramics measured from 1 KHz to 1 MHz are shown in Fig. 3(a)-3(d), respectively. All compositions show the signatures of a ferroelectric transition that follows a relaxor transition upon cooling (similar phenomenon occurs on heating). The abnormal relaxor transition at the lower temperature $T_m$ is the so-called re-entrant relaxor transition. One important difference with the two dielectric anomalies is that the lower-in-temperature one around 680 K ($T_{RR}$) shows a strong frequency dispersion, while the higher-in-temperature one is a typical DPT. Here for 0.6Bi$_{(1-x)}$(Mg$_{1/2}$Ti$_{1/2}$)O$_3$-0.4PbTiO$_3$, comparing to a typical re-entrant relaxor transition, no broad permittivity plateau is found because the $T_m$ and $T_{RR}$ is very close. It suggests that the *P4mm* ferroelectric microdomains remains in the re-entrant relaxor and the frozen PNRs make a negative influence on the switching/growth of the microdomains. The higher-in-temperature DPT around 900 K is associated with the ferroelectric to paraelectric phase transition rather than the effect of space charges.[17]

In order to gain further insights into the re-entrant relaxors, we employ the macroscopic and phenomenological approach[18,19] to describe and fit the $\varepsilon'$ of 0.6Bi$_{(1-x)}$(Mg$_{1/2}$Ti$_{1/2}$)O$_3$-0.4PbTiO$_3$. In this approach, the dielectric permittivity is proposed to following expressions:[12,20]

$$\varepsilon(T)_{reentrance} = \frac{\varepsilon_{11}}{1+b_1 \exp(-\frac{\theta_1}{T})} P_{11}(E_{b1},T) + \varepsilon_{12}(E_{b1},T) \qquad (1)$$

$$\varepsilon(T)_{DPT} = \frac{\varepsilon_{21}}{1+b_2 \exp(-\frac{\theta_2}{T})} P_{21}(E_{b2},T) + \varepsilon_{22}(E_{b2},T) \qquad (2)$$

$$\varepsilon(T) = \varepsilon(T)_{reentrance} + \varepsilon(T)_{DPT} \qquad (3)$$

where $\varepsilon_{11}$, $\varepsilon_{12}$, $\varepsilon_{21}$, $\varepsilon_{22}$, $E_{b1}$, $E_{b2}$, $\theta_1$, and $\theta_2$ are constants at a given frequency. $E_{b1}$ and $E_{b2}$ are effective potentials of PNRs corresponding to re-entrant relaxor and DPT, which can be understood as activation energy to a certain extent. The insets of Fig. 3 show the fitting curves are in very good agreement to the experimental data, indicating that the model can accurately describe the re-entrant dipole glass-like behavior and the DPT behavior. The $E_{b1}$ of the re-entrant dipole glass-like behavior and frequency has an opposite trend and the average value of $E_{b1}$ is in the range of 0.05(1)-0.07(5) eV, while the $E_{b2}$ of the DPT behavior is a constant value of 1.30(4)-1.31(3) eV (Table 2).

Remarkably, we are able to estimate the size of dipole clusters (or PNRs) in the relaxor and DPT regions using $E_{b1}$ and $E_{b2}$. The size of the PNRs is such that the direction of polarization is reoriented by thermal fluctuations of the lattice. This is consistent with the concept of a ferroelectric order crystal which includes features, such as lattice distortion, ionic disorder and charge mismatch, which limit the coherence of the spontaneous polarization. The density of the features determines the size of PNRs (coherently polarizing volume), $\lambda$. In the simplest case, each initially isotropic region is subject to distortion, on cooling through Burns temperature $T_B$, to a ferroelectric order characterized by a local, temperature-dependent polarization, $P_s$. The reorientation of the polarization vector of a single PNRs, between the variants allowed by the crystal symmetry, is considered to be a thermally activated process with activation energy $E_b$. In terms of an energy density, $\Delta G$, the activation energy is dependent on the PNRs size ($E_b=\Delta G\lambda^3$).[21] Therefore, $E_{b1}$ or $E_{b2}$ can be expressed as $\Delta G*V$, where $V$ is the volume of a single PNRs. The Landau-Devonshire formalism of free energy,[1, 12] which shows that the typical values of $\Delta G$ is in the range of $10^5 \sim 10^7$ J·m$^{-3}$. the volume of a PNRs of re-entrant relaxor and PNRs of DPT is calculated to be in the range of $(0.16 \sim 1.20) \times 10^{-27}$ m$^3$ and $(20.86 \sim 21.00) \times 10^{-27}$ m$^3$, respectively, corresponding to a lateral size in the range of 0.93~1.06 nm and 2.75~2.76 nm, respectively. The size of PNRs of re-entrant relaxor and DPT depends on the stoichiometry of bismuth due to the off-center displacement of Bi$^{3+}$ ion with lone pair of electrons creating local polarization.[22] Comparing to Bi-deficient sample, the PNRs in both the relaxor and DPT regions are larger for the Bi-excess samples, suggesting that the size of PNRs can be modified by the concentration of bismuth.

The function $\omega_1(T)$ describes the ability of dipoles to overcome the potential wells at different temperatures, as shown in Fig. 4(a), which is similar to the Fermi-Dirac function. It is close to one at low temperature while it is close to zero at high temperature.[8] The function $\omega_1(T)$ shows frequency dispersion, suggesting multiple relaxation time of dipoles in the system. Different from the function $\omega_1(T)$, the function $\omega_2(T)$ associated with DPT behavior shows a rapid decrease without frequency dispersion near the Curie temperature as shown in Fig. 4(b). It suggests that the appearance/disappearance of ferroelectric domain is very different from that of polar clusters. The phase transition of the pseudo-cubic and the tetragonal symmetry to the cubic symmetry undergoes a thermal evolution giving rise to a sharp dielectric maximum.

The empirical Vogel-Fulcher law can be employed to relate the probing frequency $f_0$ and the temperature corresponding to the maximum of the $\varepsilon'$ ($T_{RR}$) for the re-entrant dipole glass-like behavior and as shown in Fig. 5(a)-5(d).[23] The equation is given by:

$$f = f_0 \exp(-E_a / \kappa_B (T_m - T_f)) \tag{4}$$

where $E_a$ is the activation energy, $f_0$ is the Debye frequency, which is the frequency of tries to conquer the potential barrier $E_a$, $k_B$ is the Boltzmann constant ($8.617 \times 10^{-5}$ eV/K), and $T_f$ is the freezing temperature of dipoles. The fitting parameters of $E_a$, $T_f$, and $f_0$ are cataloged in Table 3. The $f_0$ for all samples are on the order of $10^{12}$ Hz, close to the typical lattice frequencies (near THz).[24] The $E_a$ is from 0.03 eV-0.08 eV depending on the Bi stoichiometry, which falls in the activation energy values of re-entrant relaxors (0.2 eV-0.3 eV) and the canonical ferroelectric relaxors (0.01 eV-0.05 eV), suggesting a strong interaction between PNRs that gives rise to the similarity to canonical ferroelectric relaxors. The increase of Bi concentration of Bi results in a decrease of $T_f$ and an increase of $E_a$, which is closely related to the fact that dipoles (and therefore polarization) arise due to the Bi off-center displacements. More Bi makes the creation of local polarization or PNRs easier, the higher $E_a$ reveals a weaker interaction between PNRs in comparison to the increased interaction inside a PNRs.

The temperature dependence of the *P-E* loop can directly show the relaxors behavior in ferroelectric materials. Polarization-electric field (*E*) and polarization current density(*J*) of the 0.6B$_{(1-x)}$MT-0.4PT measured at the temperature below $T_f$ are shown in Fig. 6. Similar to the canonical relaxor ferroelectrics, all *P-E* loops exhibit a profile between the relaxor ferroelectric (thin *P-E* loop) and the normal ferroelectric (square *P-E*) independent of temperature. It suggests that *P4mm* ferroelectric microdomains are retained in the re-entrant relaxor and the frozen PNRs make a pinning effect on *P*4*mm* ferroelectric microdomains. However, the temperature evolution of $P_{max}$ increases linearly with the increasing temperature. The increase of $P_{max}$ is attributed to the thermal activated PNRs as the $P_r$ is independent of temperature.[14,25] It suggests that the configuration of tetragonal ferroelectric microdomains remains unchanged in this temperature region (below $T_f$) since the frozen PNRs prevent the switching of tetragonal domains under electric field.[25] It is interesting that both Bi- deficient and Bi-excess sample show higher $P_{max}$, the former could be associated with bismuth vacancies and oxygen vacancies, which make the PNRs susceptible to reorientation under electric filed; while the lattice is attributed to low fraction of tetragonal phase (the PNRs contributes to the $P_{max}$ as discussed above).[26]

To clearly describe the re-entrant relaxor behavior, impedance spectroscopy with a broader and more continuous frequency spectrum was employed for the electric response of 0.6B$_{(1-x)}$MT-0.4PT ceramics. The temperature evolution of the imaginary part of the electrical modulus (*M″*) are calculated for the 60B$_{(1-x)}$MT-40PT ceramics and depicted in Fig. 7(a)-7(d). With increasing temperature, all *M″* peaks rapidly shift to higher frequencies, which exhibit a thermally activated behavior. The electric modulus physically corresponds to the relaxation of the electric field in the materials while the electric displacement remains constant, so that the electric modulus represents the real dielectric relaxation process.[27] To understand such electric modulus via relaxation mechanism, the Fourier transform of a relaxation function ($\varphi(t)$) was employed to describe:[28, 29]

$$M^* = M_\infty [1 - \int_0^\infty \exp(-\omega t)(-\frac{d\phi}{dt})dt] \qquad (5)$$

The function $\varphi(t)$ is modified by the function of Kohlrausch-Williams-Watts (KWW), the imaginary part of the electric modulus can be described:

$$M'' = M''_{max} / \{(1-\beta) + \frac{\beta}{1+\beta}[\beta(\omega_{max}/\omega) + (\omega/\omega_{max})^\beta]\} \tag{6}$$

where $M''_{max}$ is the peak value of $M''$ and $\omega_{max}$ is the corresponding frequency of $M''_{max}$. The index $\beta$ (0.1] indicating a deviation of dielectric relaxation from ideal Debye relaxation. Increasing the value of $\beta$, the relaxations approach the ideal Debye relaxation. In an ideal Debye medium, molecules are independent, there are no interactions among them, and the dipole-moment of the molecules keeps constant. The molecules will only freeze at 0 K.[12] The material under investigation here is certainly not an ideal Debye medium and the deviation can be deduced from the value of $\beta$. The temperature dependence of $\beta$ in $0.6B_{(1-x)}MT-0.4PT$ ceramics is shown in Fig. 8(a)-8(d). The changes can be understood with the following process.

The sketch of the dynamics of two kinds of PNRs of a re-entrant relaxor is shown in Fig 9. From very high temperature, when the temperature decreases the lattice is slightly distorted and ions displaced, and as the interaction between dipoles are enhanced, PNRs appear at $T_{B1}$. The dynamic of PNRs and phase transition can be determined according to the value of $\beta$. Initially, PNRs with *P4mm* symmetry emerge at high temperature ($T_{B1}$), and the interaction between them increases with decreasing temperature. The DPT behavior of $0.6B_{(1-x)}MT-0.4PT$ appears at near 900 K ($T_m$) when the *P4mm* PNRs collaborate and become *P4mm* microdomains. Further decreasing the temperature, the freezing of PNRs and the growth of microdomains weaken the interaction between them, therefore, $\beta$ increases. However, what makes re-entrant relaxors so special is that, at about 800 K, a new type of PNRs (rhombohedral symmetry) appears in the tetragonal (or the residual cubic) matrix at $T_{B2}$, which leads to some new dynamics and a new frequency dispersion for the temperature dependent dielectric permittivity, resulting in the re-entrant relaxor behavior. The dielectric anomaly near 680 K ($T_{RR}$) shows a strong frequency dispersion, suggesting the formation of re-entrant relaxor. The new PNRs grow up with the decrease of temperature and its size reaches maximum at $T_f$ when the interaction between PNRs is the strongest and $\beta$ has the smallest. Finally, the new PNRs are frozen below $T_f$. Below $T_f$, the temperature dependence of $P_r$ is attributed to the $P4mm$ ferroelectric microdomains and frozen PNRs. Meanwhile, $P_r$ slightly increases with temperature because of the re-orientation of the thermally activated PNRs (Fig. 6). On the other hand, the interaction between them become weak with decreasing temperature, therefore, $\beta$ increases monotonously with decreasing temperature. We note, however, that the mechanism for the appearance of PNRs is different from those appearing in the $Pb(Mg_{1/3}Nb_{2/3})O_3$-$PbTiO_3$.[18]

It is worth noting that for $0.6B_{(1-x)}MT$-$0.4PT$ the temperature dependence of the relaxation time does follow the Vogel-Fulcher law below $T_m$, which suggests the relaxation time of dipole clusters or PNRs increases during the cooling process and hence the correlation among them increases.[30] Comparing the temperature evolution of $M''$ to temperature dependence of $\varepsilon'$, it can be seen that the critical temperatures corresponding to the freezing and formation of PNRs in the polar phase (re-entrant relaxor) as well as the DPT phase.

**4 Conclusions**

The complex perovskite oxide $0.6B_{(1-x)}MT$-$0.4PT$ near MPB between the tetragonal and the pseudo-cubic phase was prepared. Re-entrant dipole glass behavior of $0.6B_{(1-x)}MT$-$0.4PT$ ceramics was characterized by the Vogel-Fulcher law and the phenomenological statistical model. As the key feature of relaxors, the size of PNRs in paraelectric and ferroelectric states is found to be in the range of 0.93~1.06 nm and 2.75~2.76 nm, forming the re-entrant relaxor transition and the DPT, respectively. Furthermore, it is found that three critical temperatures can be seen both in $\beta$ and in relaxation time, associated with the freezing and formation of PNRs, which is affected by the stoichiometry of bismuth. This investigation makes the dynamic behavior of PNRs clear and the provides insights as modifying the re-entrant relaxor behaviors.

## Acknowledgments

This work was financially supported by the Natural Science Foundation of China (Grant Nos. 11564010, 11974268), the Natural Science Foundation of Guangxi (Grant Nos. AA138162, GA245006, FA198015, AA294014, BA297029 and BA245069), and High Level Innovation Team and Outstanding Scholar Program of Guangxi Institutes.

## Data availability

Data available on request from the authors. The data that support the findings of this study are available from the corresponding author upon reasonable request.

*Table:*

Table 1. The refined cell parameters of the $0.6B_{(1-x)}MT-0.4PT$.

| 0.6BMT-0.4PT | $a=b_{P4mm}$ (Å) | $c_{P4mm}$ (Å) | $V_{P4mm}$ (Å$^3$) | Fraction of P4mm |
|---|---|---|---|---|
| -2%Bi | 3.95992(1) | 4.03915(9) | 63.26 | 60.33% |
| 0%Bi | 3.96101(6) | 4.03193(5) | 63.26 | 60.83% |
| 2%Bi | 3.96096(1) | 4.03189(9) | 63.25 | 57.59% |
| 4%Bi | 3.95984(6) | 4.03189(9) | 63.16 | 58.92% |
| 0.6BMT-0.4PT | $a=b=c_{Pm-3m}$ (Å) | | $V_{Pm-3m}$ (Å$^3$) | Fraction of Pm-3m |
| -2%Bi | 3.98263(4) | | 63.17 | 39.67% |
| 0%Bi | 3.98243(4) | | 63.16 | 39.17% |
| 2%Bi | 3.98319(1) | | 63.19 | 42.41% |
| 4%Bi | 3.98202(4) | | 63.14 | 41.08% |

Table 2. Fitting parameters of the dielectric permittivity of the $0.6B_{(1-x)}MT-0.4PT$ ceramics obtained using the macroscopic and phenomenological.

| | $0.6B_{0.98}MT-0.4PT$ | $0.6BMT-0.4PT$ | $0.6B_{1.02}MT-0.4PT$ | $0.6B_{1.04}MT-0.4PT$ |
|---|---|---|---|---|
| $\varepsilon_{11}$ | 5020.86 | 5160.16 | 4963.51 | 5593.72 |
| $E_{b1}$ | 0.05(3) | 0.05(1) | 0.05(8) | 0.07(5) |
| $\varepsilon_{21}$ | 643.00 | 444.99 | 437.40 | 887.35 |
| $b_1$ | 6.912 x 10$^6$ | 7.061 x 10$^6$ | 2.396 x 10$^7$ | 1.113 x 10$^8$ |
| $\theta_1$ | 13028.97 | 13038.96 | 13919.18 | 15047.82 |
| $\varepsilon_{21}$ | 2.922 x 10$^{10}$ | 2.932 x 10$^{10}$ | 2.721 x 10$^{10}$ | 2.976 x 10$^{10}$ |
| $E_{b2}$ | 1.31(3) | 1.30(9) | 1.30(4) | 1.30(7) |
| $\varepsilon_{22}$ | -1302.44 | -1233.87 | -850.67 | -868.94 |
| $b_2$ | 2.741 x 10$^{10}$ | 2.883 x 10$^{10}$ | 3.659 x 10$^{10}$ | 2.957 x 10$^{10}$ |
| $\theta_2$ | 21639.91 | 21702.80 | 21856.82 | 21582.58 |

Table 3. Fitting parameters for different compositions of $0.6B_{(1-x)}MT-0.4PT$ ceramics obtained using the Vogel-Fulcher.

|  | $T_f$ (K) | $f_0$ (Hz) | $E_a$ (eV) | *Ref.* |
|---|---|---|---|---|
| $0.6B_{0.98}MT-0.4PT$ | 691.48 | $2.58 \times 10^{12}$ | 0.03 |  |
| $0.6BMT-0.4PT$ | 665.67 | $2.65 \times 10^{12}$ | 0.04 |  |
| $0.6B_{1.02}MT-0.4PT$ | 655.30 | $2.98 \times 10^{12}$ | 0.05 | *This work* |
| $0.6B_{1.04}MT-0.4PT$ | 645.14 | $4.39 \times 10^{12}$ | 0.08 |  |
| 95BT-5BS | 222.21 | $2.31 \times 10^{7}$ | 0.02 | [5] |
| 50KNN-50BNT | 214.63 | $5.41 \times 10^{13}$ | 0.20 | [13] |
| 35.5BS-62PT-0.025PCN | 630.25 | $1.79 \times 10^{13}$ | 0.05 | [14] |
| PMN-10PT | 296.13 | $2.4 \times 10^{12}$ | 0.04 | [8,18] |

Table 4. The freezing temperature of dipoles ($T_f$), and the temperature of diffuse phase transition ($T_B$) of the $0.6B_{(1-x)}MT-0.4PT$ ceramics obtained using the deviation of dielectric relaxation from ideal Debye relaxation.

|  | $0.6B_{0.98}MT-0.4PT$ (K) | $0.6BMT-0.4PT$ (K) | $0.6B_{1.02}MT-0.4PT$ (K) | $0.6B_{1.04}MT-0.4PT$ (K) |
|---|---|---|---|---|
| $T_{f(V-F)}$ | 691.48 | 665.67 | 655.30 | 645.14 |
| $T_f$ | 689.67 | 664.15 | 653.59 | 643.55 |
| $T_{B(\varepsilon')}$ | 910.37 | 904.50 | 895.18 | 890.25 |
| $T_B$ | 904.29 | 900.52 | 890.36 | 888.85 |

*Figure captions*:

**Fig. 1.** Rietveld refined x-ray diffraction patterns using *P4mm+Pm-3m* space groups of (a) $0.6B_{0.98}MT$-$0.4PT$, (b) $0.6BMT$-$0.4PT$, (c) $0.6B_{1.02}MT$-$0.4PT$, and (d) $0.6B_{1.04}MT$-$0.4PT$ ceramics.

**Fig. 2.** The surface SEM images of (a) $0.6B_{0.98}MT$-$0.4PT$, (b) $0.6BMT$-$0.4PT$, (c) $0.6B_{1.02}MT$-$0.4PT$, and (d) $0.6B_{1.04}MT$-$0.4PT$ ceramics. The statistic distributions of the gain size in ceramics using Gauss distribution.

**Fig. 3.** The temperature dependence of real part of dielectric constant and dielectric loss of (a) $0.6B_{0.98}MT$-$0.4PT$, (b) $0.6BMT$-$0.4PT$, (c) $0.6B_{1.02}MT$-$0.4PT$, and (d) $0.6B_{1.04}MT$-$0.4PT$ ceramics. Inset: Fittings of the dielectric permittivity of sample with phenomenological statistical model in 1 MHz.

**Fig. 4.** (a) $w_1$ vs temperature. (b) $w_2$ vs temperature in $0.6B_{(1-x)}MT$-$0.4PT$ ceramics.

**Fig. 5.** Plots of ln$\omega$ *vs.* $T_m^{'}$. Solid curves are fitted according to the Vogel-Fulcher relation of (a) $0.6B_{0.98}MT$-$0.4PT$, (b) $0.6BMT$-$0.4PT$, (c) $0.6B_{1.02}MT$-$0.4PT$, and (d) $0.6B_{1.04}MT$-$0.4PT$ ceramics.

**Fig. 6.** Polarization-electric filed (*P-E*) at different temperatures. Inset: $P_r$ and $P_{max}$ vs temperature of (a) $0.6B_{0.98}MT$-$0.4PT$, (b) $0.6BMT$-$0.4PT$, (c) $0.6B_{1.02}MT$-$0.4PT$, and (d) $0.6B_{1.04}MT$-$0.4PT$ ceramics.

**Fig. 7.** Frequency dependence of the imaginary part of the electric modulus for 60BMT-40BCT measured at various temperature of (a) $0.6B_{0.98}MT$-$0.4PT$, (b) $0.6BMT$-$0.4PT$, (c) $0.6B_{1.02}MT$-$0.4PT$, and (d) $0.6B_{1.04}MT$-$0.4PT$ ceramics.

**Fig. 8.** Temperature dependence of *β* of (a) $0.6B_{0.98}MT$-$0.4PT$, (b) $0.6BMT$-$0.4PT$, (c)

0.6B$_{1.02}$MT-0.4PT, and (d) 0.6B$_{1.04}$MT-0.4PT ceramics. Inset: Temperature dependence of the main relaxation time according to the $M''$ peak. The red line is the Vogel-Fulcher fitting.

**Fig. 9.** The sketch of the dynamics of two kinds of PNRs of a re-entrant relaxor.

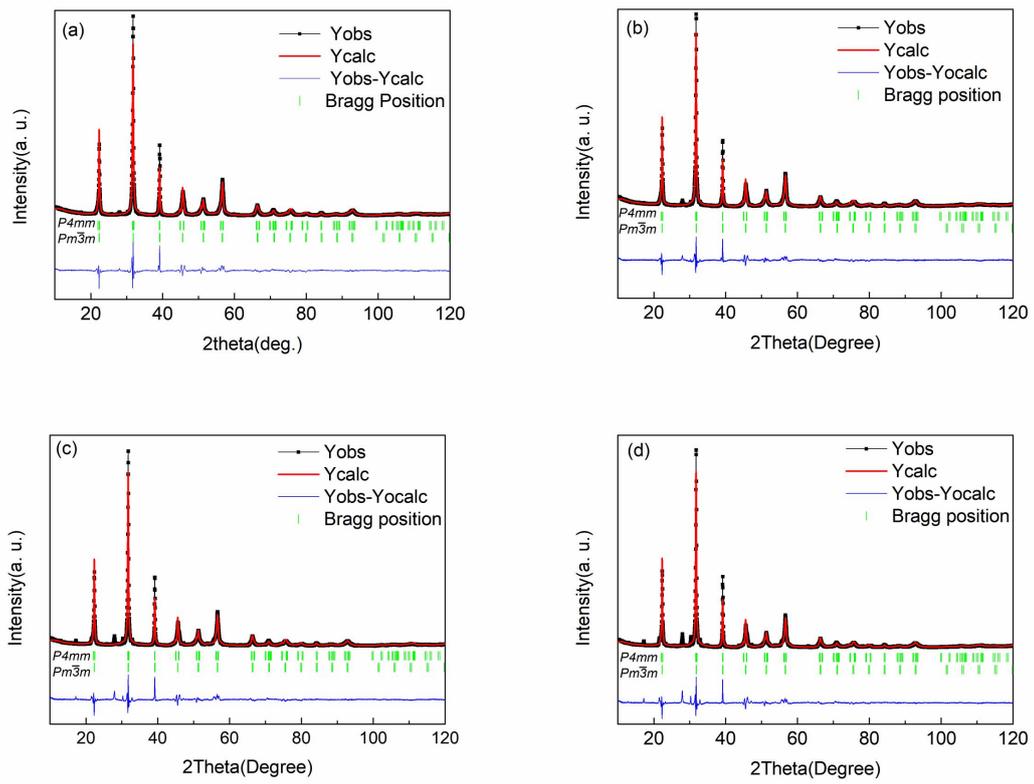

Fig. 1

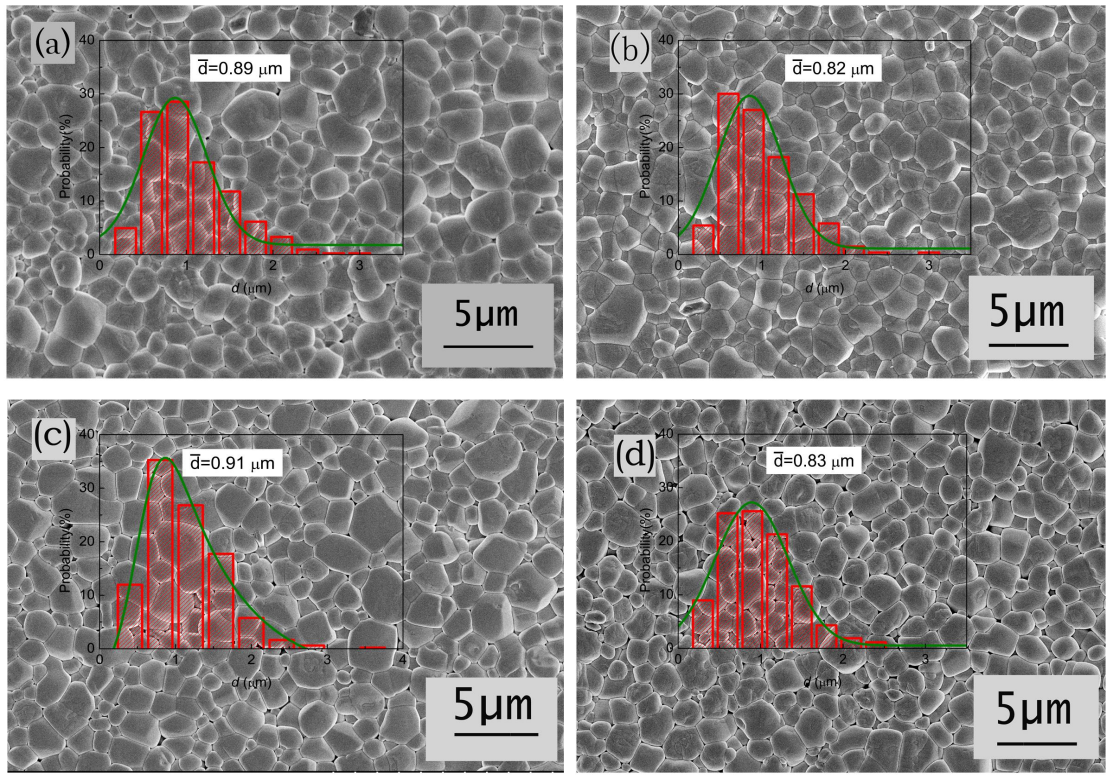

Fig. 2

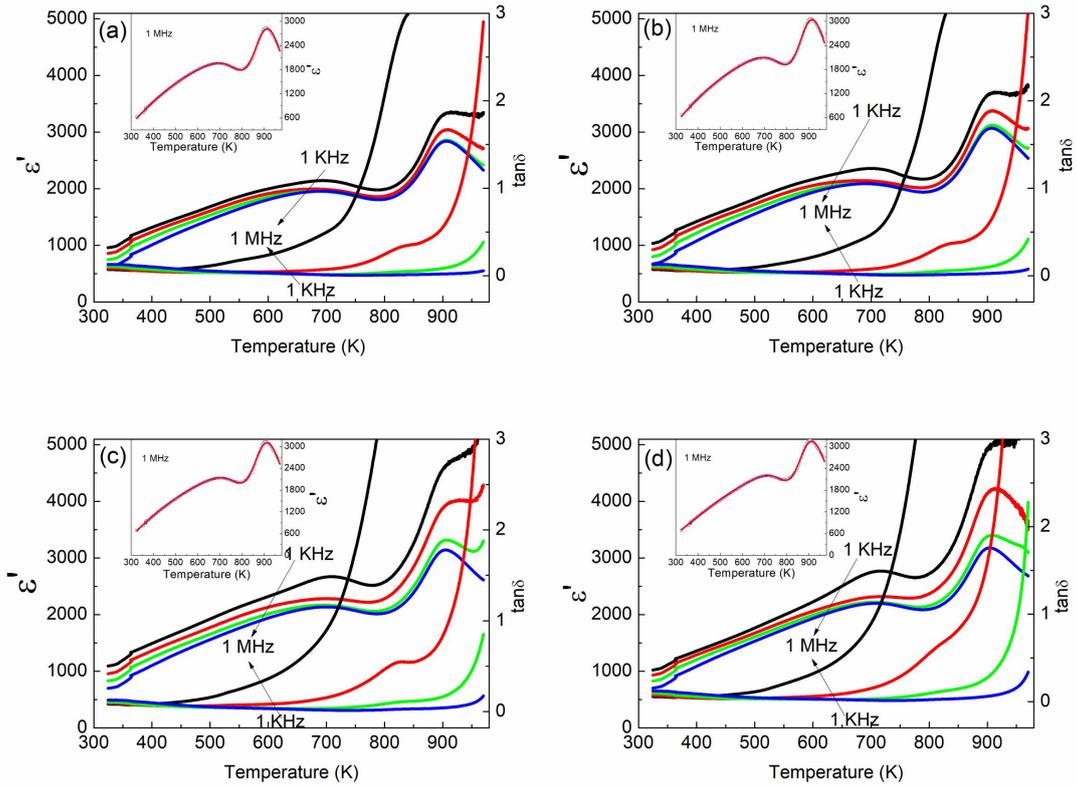

Fig. 3

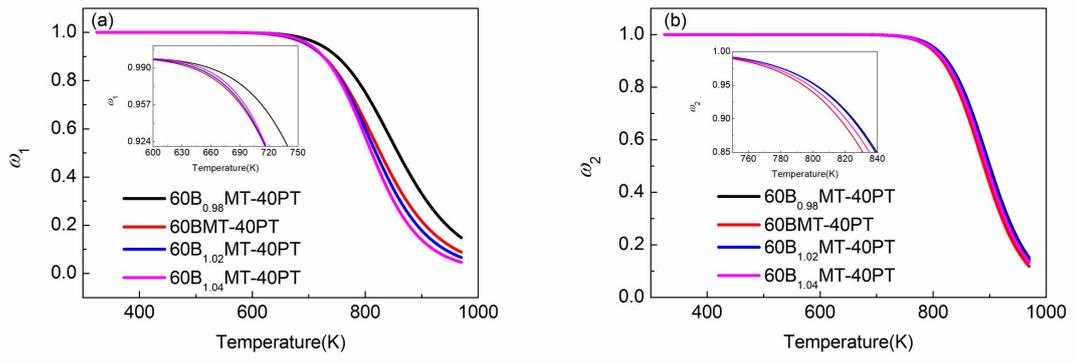

Fig. 4

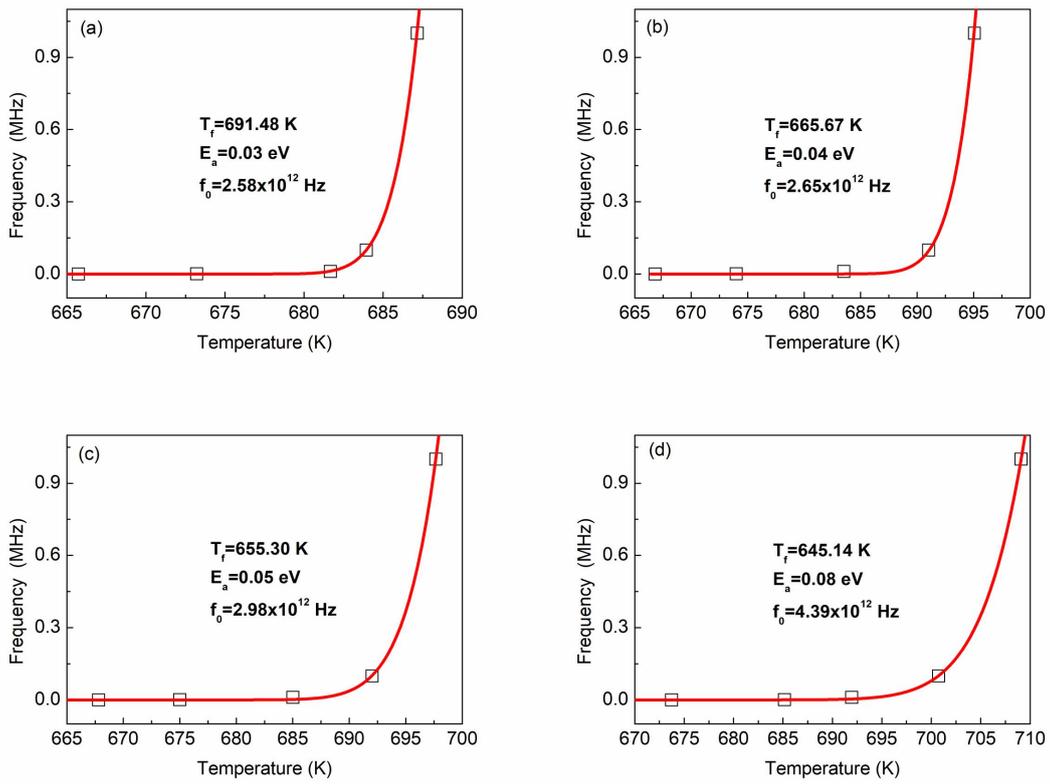

Fig. 5

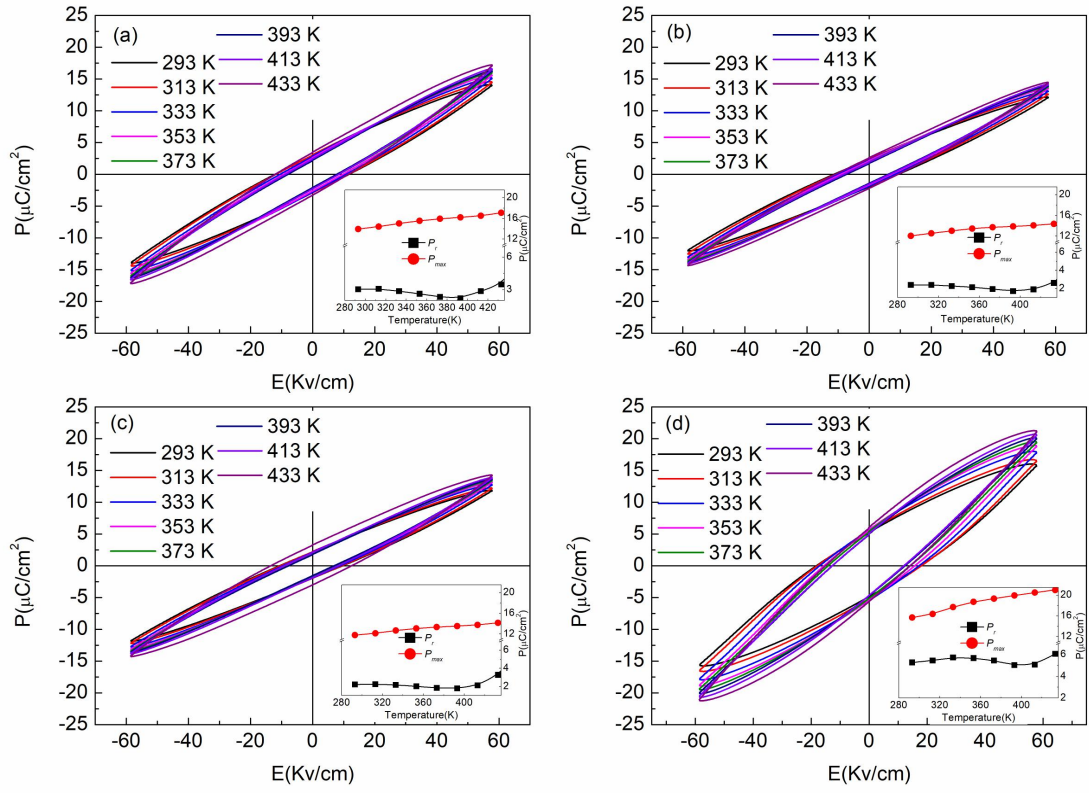

Fig. 6

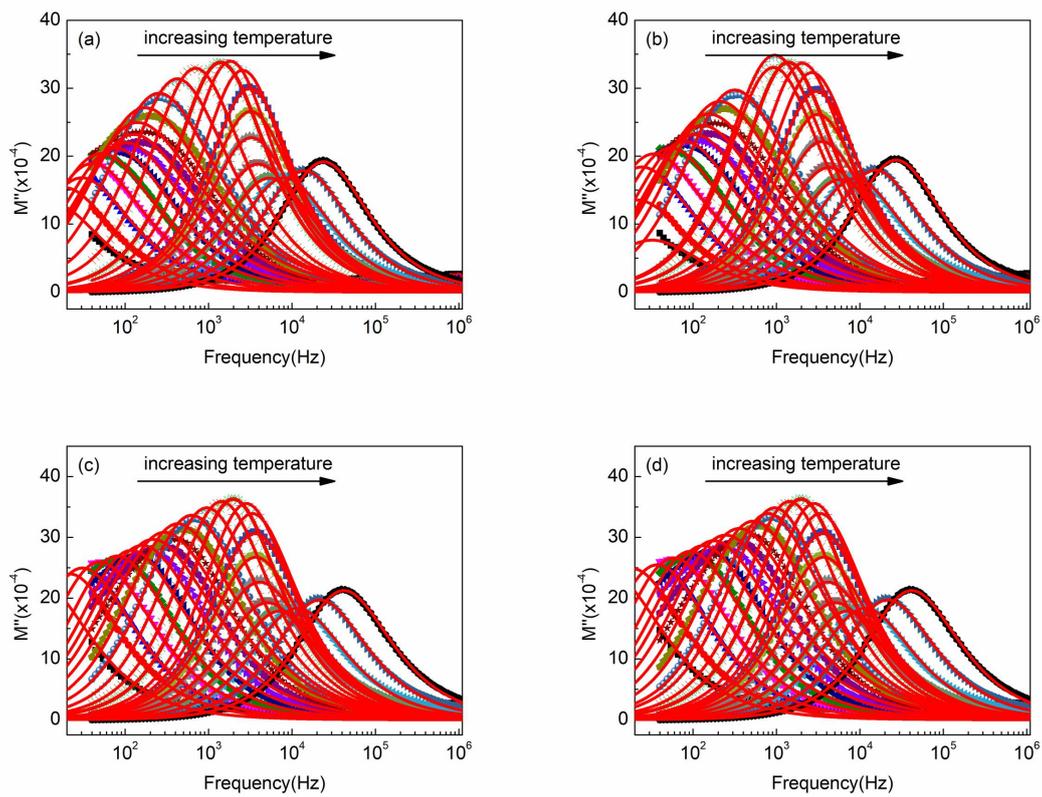

Fig. 7

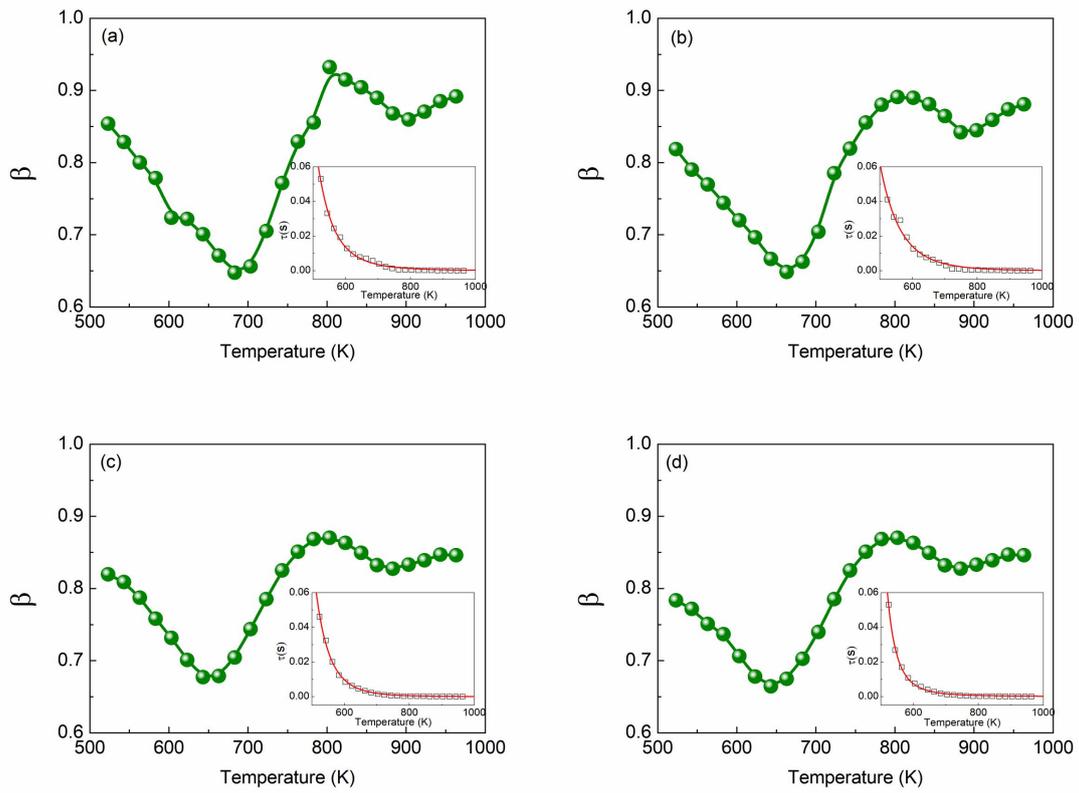

Fig. 8

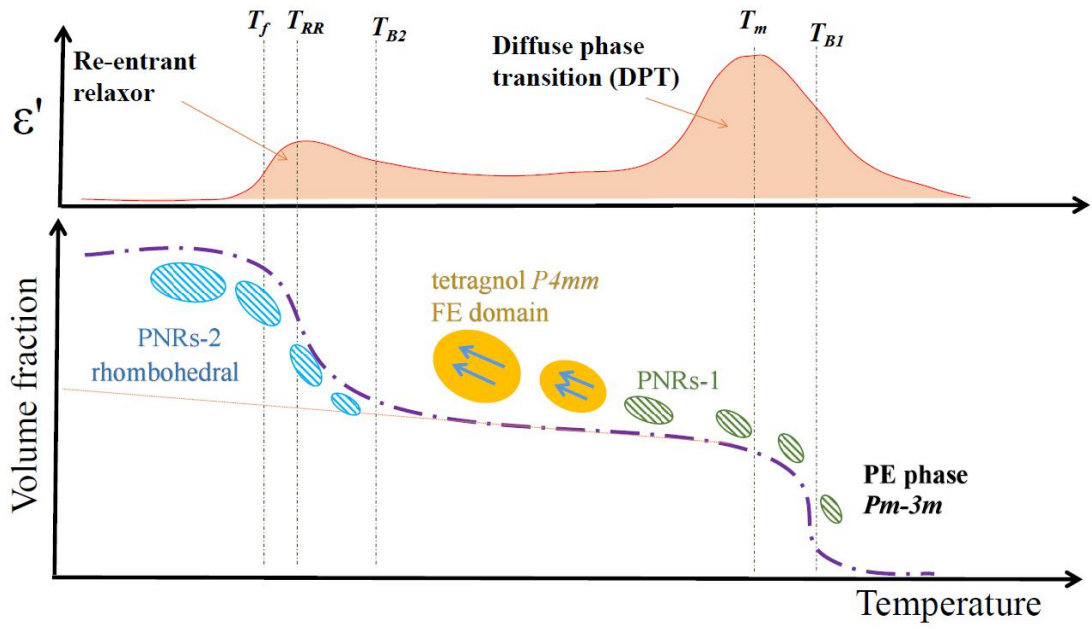

Fig. 9